\documentclass{astronomyletters}

\usepackage{mathptmx}
\usepackage{hyperref}
\usepackage{babel}
\usepackage{graphicx}
\usepackage{natbib}
\usepackage[dvipsnames]{xcolor}
\usepackage{amsmath}
\usepackage{amssymb}
\usepackage{subcaption}

\usepackage{color}

\usepackage{xspace} 
\usepackage{float}  

 \graphicspath{{./}} 
\DeclareGraphicsExtensions{.pdf,.png,.jpg,.png}

\usepackage{ulem}

\newcommand{\mlens}{{\textit{ Microlens}}}

\newcommand{\chilim}{\ensuremath{\chi_{crit}^2}\xspace}
\newcommand{\fplps}{\ensuremath{\nu_{PLPS}}\xspace}
\newcommand{\gnoplps}{\ensuremath{g_{PBH}}\xspace}

\newcommand{\plps}{PLPS\xspace}
\newcommand{\noplps}{no-PLPS\xspace}

\begin{document}

	\titles[Microlensing by Cluster of PBHs]{Microlensing by Cluster of Primordial Black Holes}
	
	\authors[Toshchenko et al.]{
		\nextauth{Toshchenko~K.A.}{1},
		\nextauth[petr.baklanov@itep.ru]{Baklanov~P.V.}{1,2,3},
		\nextauth{Belotsky~K.M.}{4,5},
		\nextauth{Blinnikov~S.I.}{1,6}
	}

	\affiliations{
		\nextaffil{The National Research Centre Kurchatov Institute, 123182, Moscow, Russia}
		\nextaffil{Space Research Institute, Russian Academy of Sciences, 117997, Moscow, Russia}
		\nextaffil{Lebedev Physical Institute, Russian Academy of Sciences, 119991, Moscow, Russia}
		\nextaffil{National Research Nuclear University (MEPhI), 115409, Moscow, Russia}
		\nextaffil{Novosibirsk State University, 630090, Novosibirsk, Russia}
		\nextaffil{Sternberg Astronomical Institute, Moscow State University, 119234, Moscow, Russia}
	}

	\def\received{Submitted ??.??.2025; revised ??.??.2025; accepted ??.??.2025}

	\def\journame{ASTRONOMY LETTERS ~ Vol. ?? ~ No.~?? ~ 2025}


	\wideabstract{
Numerous microlensing survey programs have constrained the possibility of dark matter existing in the form of compact objects within the Galactic halo. These constraints on the dark matter fraction were derived under the assumption of isolated, widely separated objects. This work investigates microlensing by primordial black holes (PBHs) organized into clusters. In this scenario, it is necessary to account for both the influence of neighboring PBHs and the collective gravitational potential of the entire cluster, which significantly complicates the microlensing light curve. Events exhibiting such complex light curves elude detection in observational experiments such as MACHO, EROS, OGLE, POINT-AGAPE, and HSC. It is demonstrated that a significant fraction of PBH dark matter (up to 93\% for the models studied) remains undetected in these observational data. However, for all considered cluster models, a substantial population of PBHs still behaves as isolated lenses. Consequently, the clustering of PBHs does not completely eliminate the microlensing constraints on the PBH contribution to dark matter.

}

\section{INTRODUCTION}
\addcontentsline{toc}{section}{INTRODUCTION}


One of the fundamental challenges in astrophysics is elucidating the nature of dark matter. A viable hypothesis to explain dark matter is the existence of primordial black holes (PBHs), proposed in the pioneering works of \citep{Zeldovich1966,hawking1971gravitationally}. However, numerous constraints have been established to date on the fraction of dark matter in the form of PBHs, defined as \(f_{PBH}=\Omega _{PBH}/\Omega _{CDM}\) \citep{Dolgov2018,carr2021constraints}. In the mass range of \(10^{-11}\div 10^{5}M_{\odot }\), the limits on \(f_{PBH}\) are imposed by observational experiments searching for Massive Astrophysical Compact Halo Objects (MACHOs) in the Galactic halo. The MACHO, EROS, and OGLE collaborations searched for MACHO signatures toward the Large and Small Magellanic Clouds (LMC and SMC); additionally, OGLE surveyed the halo toward the Galactic bulge, while the POINT-AGAPE and Subaru Hyper Suprime-Cam (HSC) collaborations conducted observations toward M31. The detected microlensing light curves appear as optical transients in the form of flaring events. Specialized algorithms are employed to distinguish these events from other transients, such as stellar variability, novae, and supernovae \citep{MACHO2000}. The concept of investigating massive objects in the Galaxy by observing the brightening of distant light sources was first proposed by \cite{Byalko1969orig}. Subsequently, \cite{Paczynski:1985jf} advanced this idea by suggesting the use of microlensing as a method to search for MACHOs. 
The first microlensing observations \citep{MACHO2000} constrained the fraction of dark matter in the form of isolated MACHOs to \(f_{PBH}\le 0.2\) for the S-model of the Halo for compact objects of \(0.5M_{\odot }\). A similar limit of \(f_{PBH}\le 0.2\) was obtained by the POINT-AGAPE collaboration for compact objects in the range \(0.5\div 1M_{\odot }\) \citep{refId0}. The EROS \citep{2007A&A...469..387T} and OGLE \citep{2019PhRvD..99h3503N,mroz2025} collaborations have almost completely ruled out MACHO objects in the range \(10^{-5}\div 10^{2}M_{\odot }\), where the upper limit is \(f_{PBH}<0.01\). Smaller masses in the range \(10^{-11}\div 10^{-6}M_{\odot }\) were excluded by the HSC collaboration, resulting in an upper limit of \(f_{PBH}<2\times 10^{-3}\) \citep{2019NatAs...3..524N}. Observations conducted with the Kepler space telescope have also strongly constrained \(f_{PBH}<0.3\) for small masses in the range \(10^{-9}\div 10^{-7}M_{\odot }\) \citep{PhysRevLett.111.181302, Griest_2014}. 

The aforementioned constraints are based on a statistical analysis of microlensing events within the framework of isolated single-lens models and, therefore, do not account for the case of clustered lenses. Individual microlensing events exhibiting light curve profiles distinct from those of single lenses were selected manually and interpreted as binary lenses. However, their sparse number prevented the determination of both the event rate and their resulting constraints on $f_{PBH}$
\citep{2000ApJ...541..270A}.

To date, an extensive set of constraints on \(f_{PBH}\) has been established \citep{Dolgov2018,carr2021constraints}. However, PBH clustering may fundamentally alter this landscape. Specifically, it has been demonstrated that PBH clusters can account for the LIGO/Virgo observational results \citep{korol2020merger,2021Univ....7...18T,2020JCAP...11..036A,Atal_2020}. In the case of clustered PBHs, gravitational microlensing not only produces effects distinct from those of isolated PBHs but also provides insights into potential clustering mechanisms. The influence of PBH clustering on microlensing has been examined in previous studies. For instance, \cite{gorton2022effect, petavc2022microlensing} utilized PBH cluster models where individual PBHs did not interfere with each other’s caustics; consequently, the resulting light curves were indistinguishable from those of a single lens. Conversely, \cite{GARCIABELLIDO2018144} assumed a PBH cluster model sufficiently compact to be treated as a single lens (with the total cluster mass). In contrast, for the PBH cluster models considered in this work, it is shown that a significant fraction of PBHs cannot be regarded as isolated lenses, and the cluster itself is not compact. In this scenario, the microlensing events yield both single-lens profiles and profiles that deviate significantly from them. 

The paper is organized as follows. First, we provide a physical description of the PBH clusters acting as gravitational lenses. This is followed by the calculation of the parameters necessary for simulating microlensing events. Next, an analysis of the modeled microlensing events is presented. Finally, the key findings are summarized in the Conclusion.

\section{PBH CLUSTERS} \label{parameters}

Two fundamentally different mechanisms for PBH cluster formation can be distinguished. In the first mechanism, PBH clusters form through the gravitational binding of PBHs generated from primordial density fluctuations. Following their formation, PBHs cluster stochastically due to their spatial distribution. Typically, Gaussian primordial curvature perturbations are considered during the inflationary stage; under certain model assumptions, these perturbations can become sufficiently large to form PBHs. The spatial distribution of these PBHs follows a Poisson distribution. Proximally located PBHs may become gravitationally bound during subsequent evolution. Such a clustering mechanism is generally weak. However, it can be enhanced by the non-Gaussianity of primordial curvature perturbations \citep{10.1093/ptep/ptz105,2015PhRvD..91l3534T,Young_2015,2019PhRvD.100l3544M,young2020} and, as recently suggested, by the presence of a scalar field in the Brans-Dicke model \citep{2020Univ....6..158B}. In this case, the PBH distribution becomes modulated.
The mass distribution of PBHs within the cluster is assumed to be either monochromatic or log-normal \citep{Dolgov1993}.

The second mechanism is based on a phase transition during inflation accompanied by the formation of domain walls \citep{2000hep.ph....5271R,2001JETP...92..921R,2005GrCo...11...99D}. A crucial aspect of this model is the presence of quantum fluctuations of the scalar field as it reaches one of its potential minima. If the field fluctuated in a certain region of space such that it shifted to a different minimum within that region (thereby generating a domain wall at the boundary), then during a subsequent e-folding near this boundary, any further fluctuations can easily push the field from one minimum to the other. Thus, if a single domain wall is initially created which later collapses into a PBH, numerous other walls (varying in size due to the different e-foldings that formed them) are generated nearby, eventually forming a PBH cluster. The mass function of PBHs within such a cluster is typically assumed to follow a falling power-law distribution \citep{Belotsky:2018wph,Khlopov:2004sc}.

\begin{figure*}[h] 
\centering
\begin{subfigure}[b]{0.49\textwidth}
	\includegraphics[width=\linewidth]{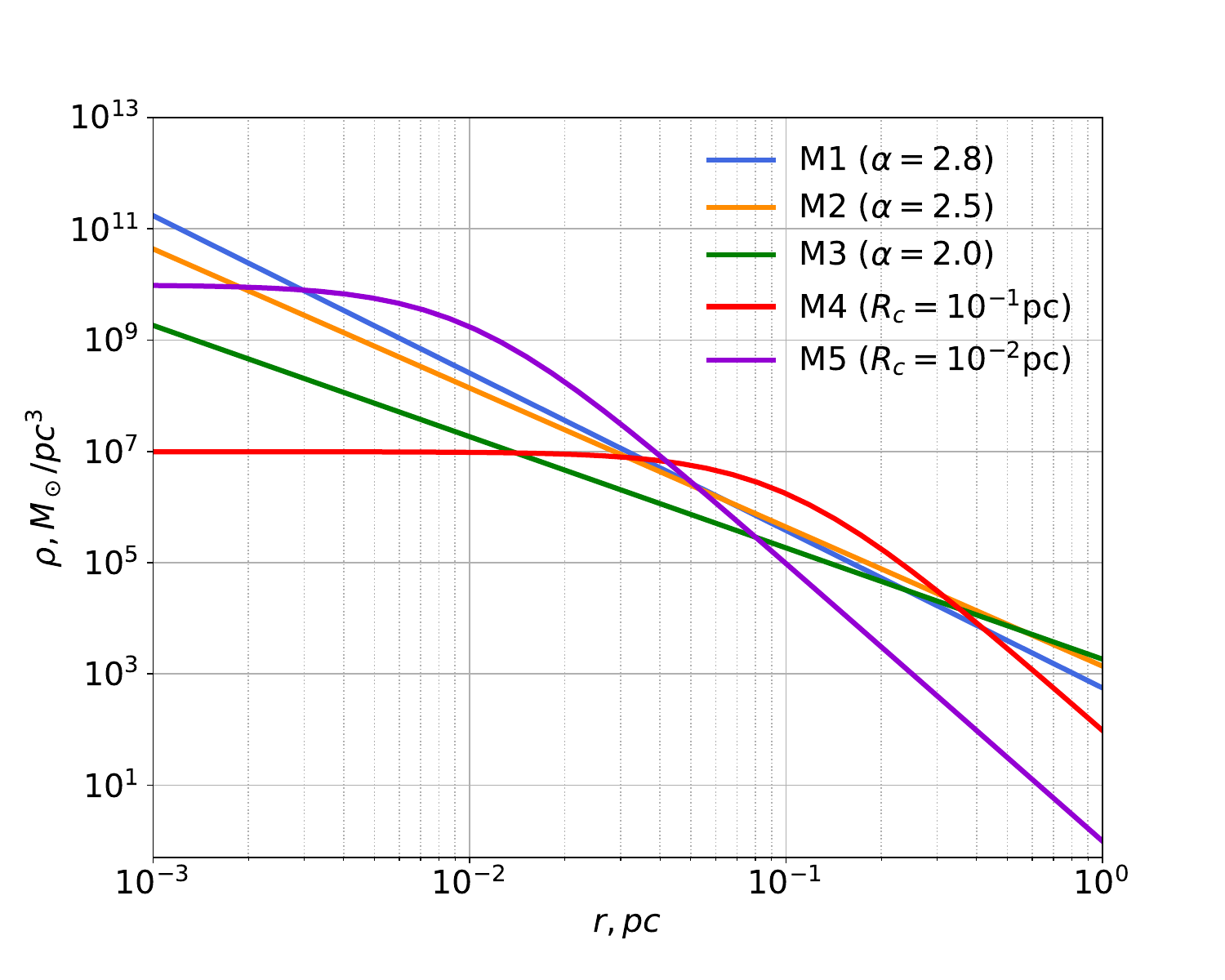}	
	\caption{Density profile for PBH cluster models
	}
	\label{Density}
\end{subfigure}
\hfill
\begin{subfigure}[b]{0.49\textwidth}
	\includegraphics[width=\linewidth]{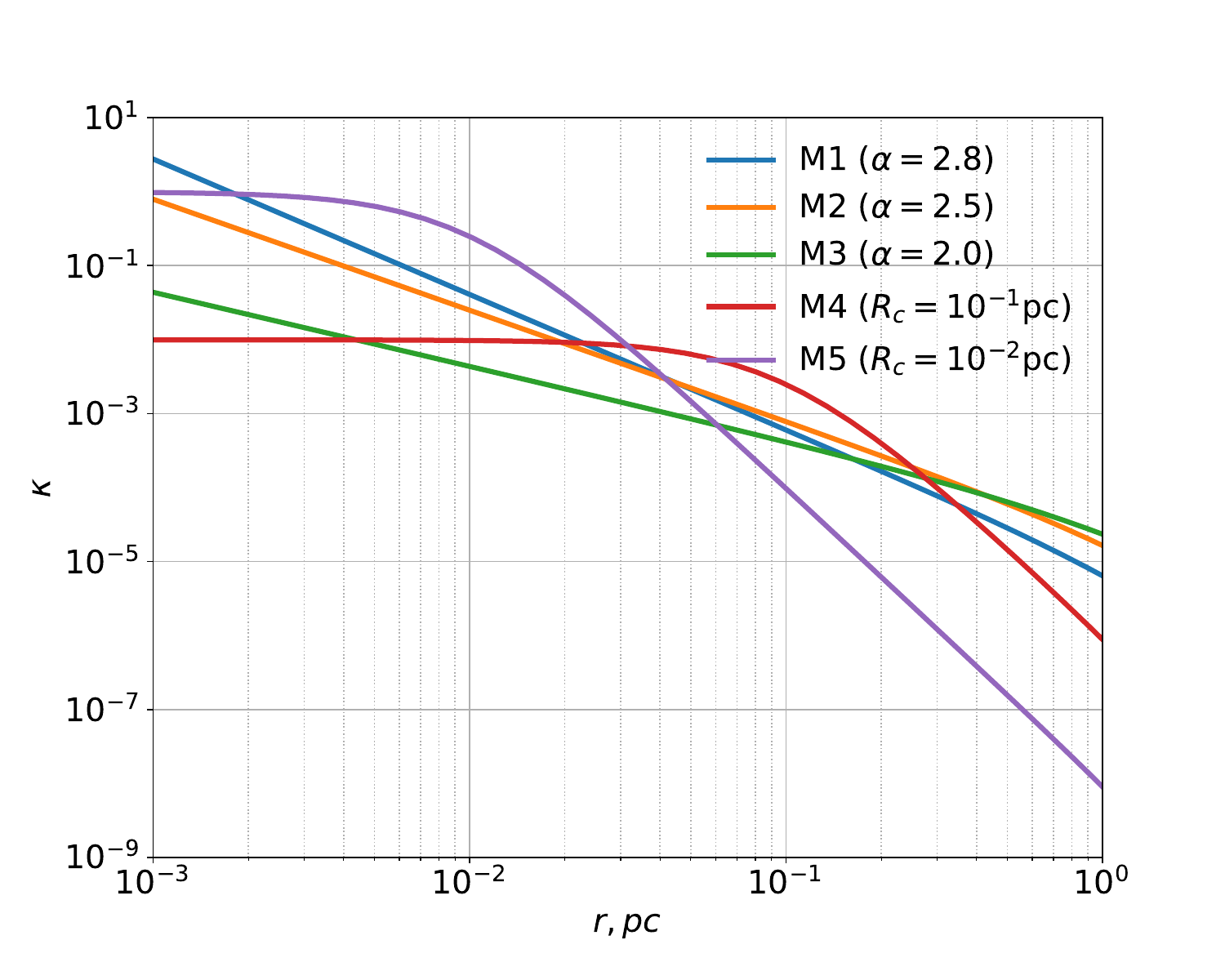}
	\caption{Optical depth $\kappa$  within the PBH cluster}		
	\label{Ka}
\end{subfigure}
\hfill
\begin{subfigure}{0.49\textwidth}
	\includegraphics[width=\linewidth]{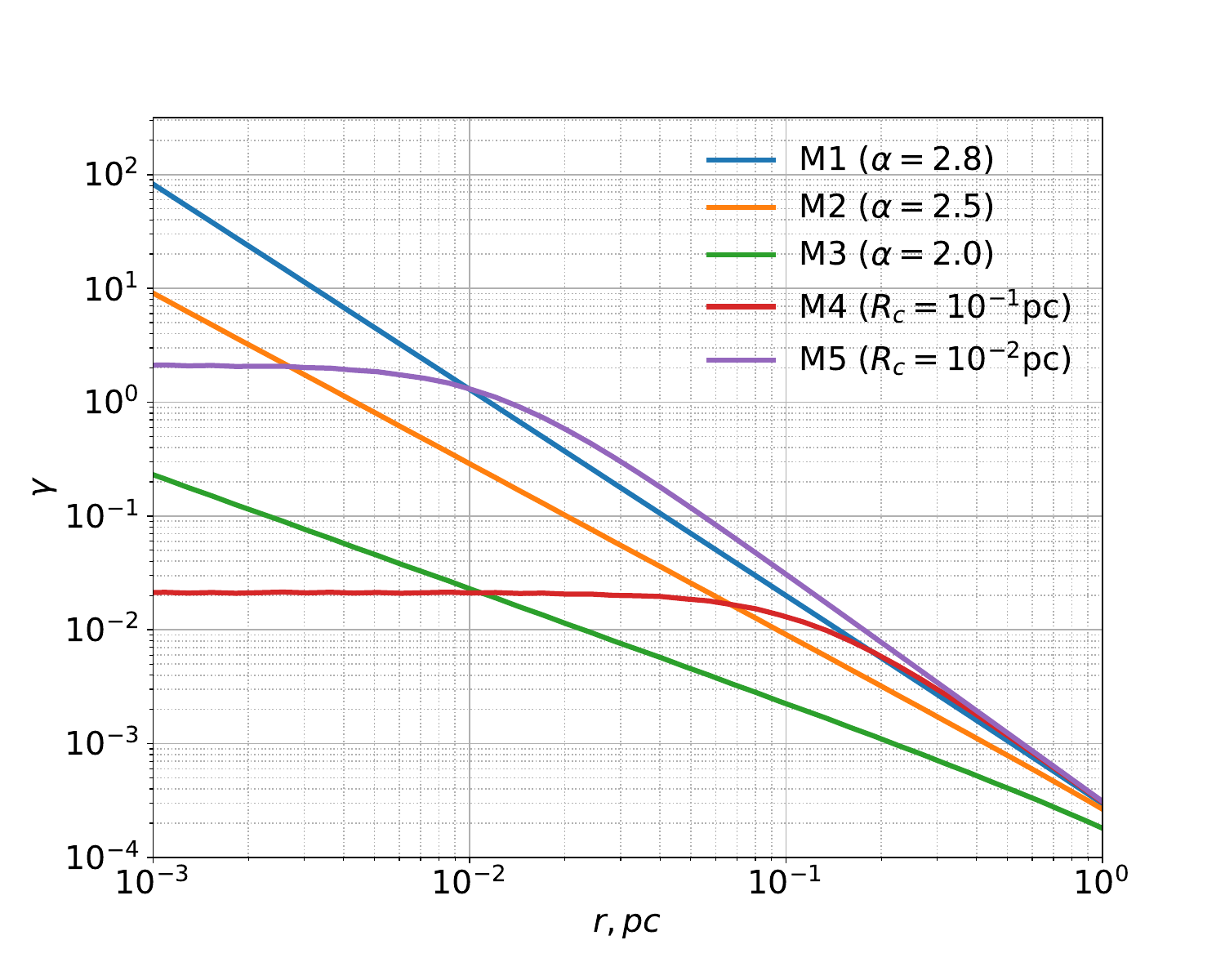}
	\caption{External shear $\gamma$ within the PBH cluster }
	\label{Ga}
\end{subfigure}
\hfill
\begin{subfigure}{0.49\textwidth}
	\includegraphics[width=\linewidth]{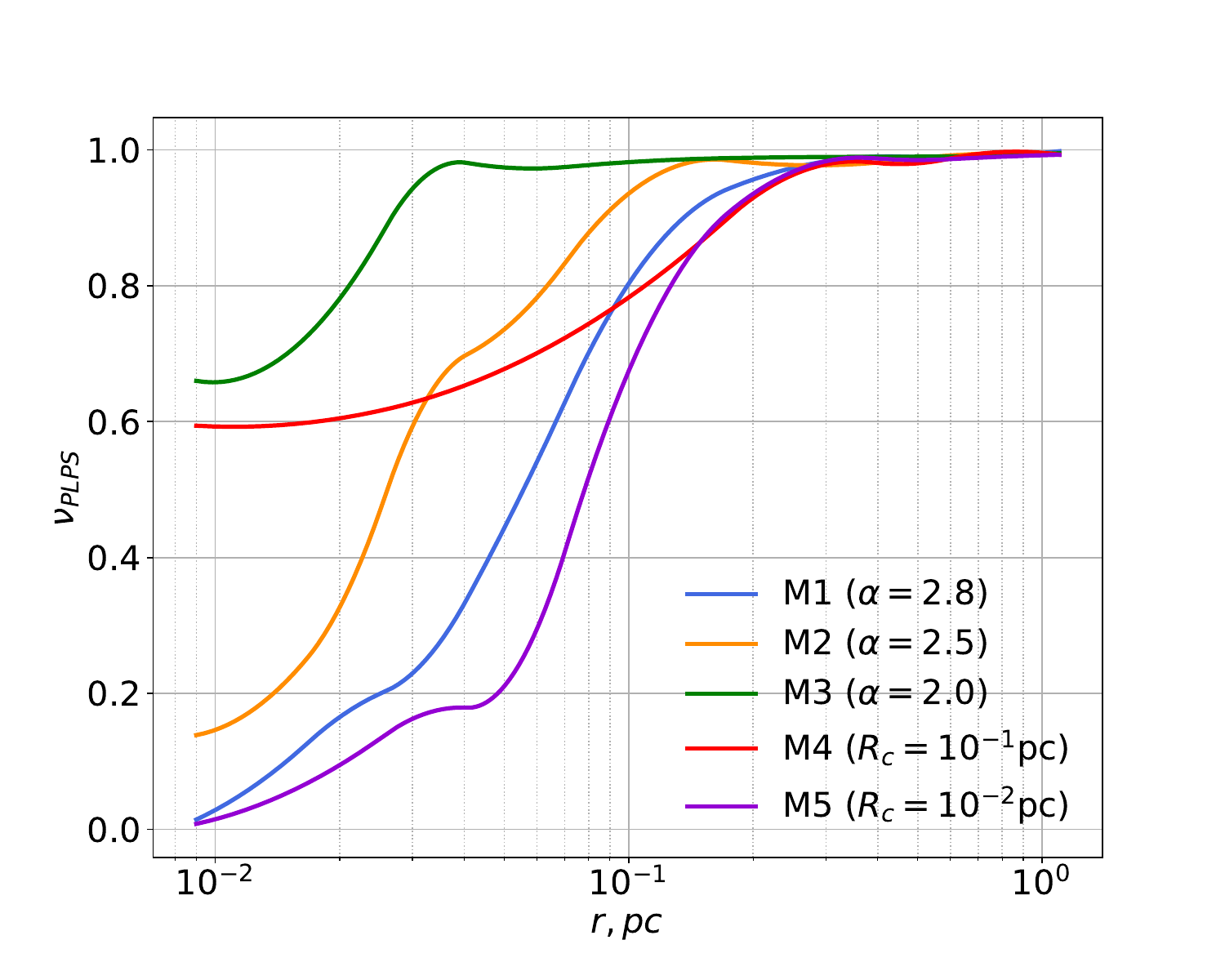}
	\caption{\plps fraction as a function of $r$ }
	\label{PercentEasy}
\end{subfigure}
\caption{Parameters of the PBH cluster models.}
\label{Pic_Parameters_Cluster}
\end{figure*}

In the second mechanism, a clustered structure for the spatial distribution of PBHs arises naturally. Therefore, in this work, the PBH cluster models were constructed based on the initial model M1, which is derived from the second mechanism and was proposed in \cite{Belotsky:2018wph}. Model M1 features a spherically symmetric density distribution, a total mass of \(M_{tot}=4\times 10^{4}M_{\odot }\), and a radius of $R_{cl} = 1$\,pc. The black hole masses in M1 range from \(10^{-4}\) to \(10^{2}M_{\odot }\), and the PBH mass function follows a power law of \(\sim M^{-2}\). The key parameters of model M1, which are inherited by all derivative PBH cluster models, are summarized in Table~\ref{tbl:params}.
\begin{table}
\caption{Common parameters for the PBH cluster models.}
	\begin{tabular}{l}
		\hline
		\hline
		Cluster radius:			$\sim 1$pc \\
		PBH mass function: 	$ \propto M^{-2} $ \\
		PBH mass range: 		$10^{-4}\div 10^{2} M_{\odot}$ \\
		Cluster total mass: 			$ 4 \times 10^4 M_\odot$ \\		
		\hline
	\end{tabular}
\label{tbl:params}
\end{table}

M1 describes a primordial PBH cluster in the early Universe and is characterized by a power-law density profile \(\rho (r)\sim r^{-\alpha }\). In \cite{Belotsky:2018wph}, the value \(\alpha =2.8\) was adopted for M1, though the formation of PBH clusters with shallower density profiles is also permitted. To investigate how the shape of the cluster density profile \(\rho (r)\) influences the microlensing effect, we consider models with flatter profiles, specifically \(\rho (r)\sim r^{-2.5}\) and \(\rho (r)\sim r^{-2}\).

Plummer models with two distinct core radii, $R_{c} = 0.01$\,pc and $R_{c} = 0.1$\,pc, were employed. This allowed for an analysis of the core radius influence on the observed microlensing effects. The designations and characteristics of the PBH cluster models are summarized in Table~\ref{tbl:models}. Figure \ref{Density} illustrates the density profile distributions for the considered PBH cluster models.

\begin{table*}[t]
\begin{center}
		\begin{tabular}{ c c c c c c c }
			\hline
			\hline
			Model & Profile $\rho(R)$ &  $l_{e}$, pc & $\theta_{e}$, $^{\prime\prime}$& $t_{e}$, year & $M_{\kappa>1}$, $M_\odot$\\
			\hline
			M1 & Power-law  $\alpha = 2.8$ & $1.7\times 10^{-3}$ & $1.4\times 10^{-2}$ & $8.3$ & $1.3\times 10^{4}$\\
			M2 & Power-law  $\alpha = 2.5$ & $8.5\times 10^{-4}$ & $7.0\times 10^{-3}$ & $4.2$ & $1.2\times 10^{3}$\\
			M3 & Power-law  $\alpha = 2.0$ & $4.0\times 10^{-5}$ & $3.3\times 10^{-4}$ & $0.2$ & $1.5\times 10^{0}$\\
			M4 & Plummer  $ R_{c} = 10^{-1} $ pc  & --- & --- & --- & --- \\		
			M5 & Plummer  $ R_{c} = 10^{-2} $ pc  & $1.0\times 10^{-3}$& $8.3\times 10^{-3}$ & $4.9$ & $4.1\times 10^{2}$\\
			\hline	
		\end{tabular}
	\end{center}
	\caption{	 	
		Model parameters for PBH clusters. All models share the same total mass, \(M_{tot}=4\times 10^{4}\,M_{\odot }\). Here, \(\rho (R)\) denotes the density profile, while \(l_{e}\) and \(\theta _{e}\) represent the linear and angular scales, respectively, of the region where the strong gravitational lensing effect occurs. The parameter \(t_{e}\) is the characteristic microlensing timescale for the PBH cluster acting as a single lens. \(M_{\kappa >1}\) is the mass of the cluster fraction responsible for strong lensing. For model M4, there is no region where \(\Sigma >\Sigma _{cr}\); consequently, M4 does not produce strong lensing effects. 	 	}
	\label{tbl:models}
\end{table*}

\section{MICROLENSING}\label{microIntro}

Observational microlensing experiments searched for brightening events of distant sources described by the Point-Lens Point-Source (\plps) model \citep{mroz2025}. In the \plps model, the magnification \(\mu \) and the light curve shape \(u(t)\) of the source are given by the following equations:
\begin{align} 
	\label{eq:One_lens_magnification}
	\mu = \frac{u^2 + 2}{u \sqrt{u^2 + 4}},  u (t)= \sqrt{u_{min}^2 + \frac{(t - t_{max})^2}{t_e^2}}
\end{align},
where \(u_{min}\) is the impact parameter, \(t_{e}\) is the Einstein-Chwolson radius crossing time, and \(t_{max}\) is the time of maximum magnification caused by microlensing. PBHs within a cluster exert mutual gravitational influence on one another, which complicates the light curve shape and causes it to deviate from the \plps model. We shall refer to microlensing events with light curves distinct from \plps as \noplps.

Ray-tracing algorithms are frequently employed to simulate light curves for multiple lenses experiencing mutual influence. We utilized an algorithm of this type, implemented in the \mlens code \citep{WAMBSGANSS1999353}, to simulate microlensing by PBH clusters. The \mlens code is highly optimized, allowing for the rapid construction of magnification maps for a large number of point lenses (up to \(10^{7}\)).

The \mlens code takes the following key parameters as input: the optical depth \(\kappa \), the external shear \(\gamma \), the mass function of the lenses, the minimum and maximum lens masses, and the size of the simulation area for which the magnification map is constructed. The optical depth \(\kappa \) for the PBH cluster was calculated according to equation (\ref{eq:formulakappa}) from \cite{Schneider1992}.
\begin{align} \label{eq:formulakappa}
	\begin{split}
		\kappa(r) &= \frac{ \Sigma(r)}{\Sigma_{cr}} \\
		\Sigma(r) &= \int^{D_s}_0 \rho(r, z) dz,  \Sigma_{cr}=\frac{c^2}{4\pi G} \frac{D_s}{D_l D_{ls}}
	\end{split}
\end{align}
where \(r\) is the distance from the lens center in the lens plane, \(z\) is the distance along the line of sight, and \(\rho \) is the mass density distribution. \(D_{s}\), \(D_{l}\), and \(D_{ls}\) are the distances from the observer to the source star, from the observer to the lens, and from the lens to the source star, respectively; \(c\) is the speed of light, and \(G\) is the gravitational constant. For the PBH cluster, the external shear \(\gamma \) was calculated according to the equation.
The external shear \(\gamma \) for the PBH cluster was calculated according to Equation (\ref{eq:formulagamma}) from \cite{Schneider1992}.
\begin{align}
	\gamma(r) = \frac{m(r)}{r^2} - \kappa(r), m(r) =  2 \int_{0}^{r} \kappa(r') r' dr'
	\label{eq:formulagamma}
\end{align}

For all investigated PBH cluster models, $\kappa$ and $\gamma$ were calculated based on their respective density profiles
The radial profiles of $\kappa (r)$ and $\gamma (r)$ for the PBH clusters are shown in Figs.~\ref{Ka} and \ref{Ga}.
Following the observations from EROS, MACHO, and OGLE, a distance of $D_s = 50$,kpc was adopted for the light curve simulations, which corresponds to the distance to the LMC. The probability of inducing a microlensing effect depends on the optical depth \(\kappa \), which, according to Eq.~(\ref{eq:formulakappa}), is maximized at $D_l = D_s/2 = 25$\,kpc.
The values of $\kappa (r)$ and $\gamma (r)$ vary significantly within the PBH cluster. The cluster was divided into a series of equal-sized areas along the radius $r$, within which $\kappa (r)$ and $\gamma (r)$ were assumed to be constant. The radial coordinates \(r\) of these areas were log-spaced from \(r_{min}\) to \(r_{max}\). The minimum radius \(r_{min}\) is constrained by the strong lensing region \(l_{e}\), which is defined by the condition \(\Sigma (l_{e})=\Sigma _{cr}\) \citep{1986MNRAS.219..333S}. 
The maximum radius \(r_{max}\) is identical for all PBH cluster models and corresponds to the M1 size of $R_{cl}=1$\,pc. The simulation area was set to \(2R_{e}(M_{\odot })\), which corresponds to a light curve duration of \(\sim 350\) days and is determined by the observational constraints of the EROS, OGLE, and MACHO experiments.
The velocity of the PBH cluster as a whole, perpendicular to the line of sight, was assumed to be the average Galactic velocity of 200 km/s. The internal proper motion of PBHs within the cluster was neglected for the following reasons. The light curves of lensing events are simulated within areas that constitute an extremely small fraction of the total cluster size --- on the order of one hundred-millionth (see Table \ref{tbl:microlens}) 
Within the simulation area, the motion of the black holes relative to the distant source can be considered rectilinear. The trajectories along which the light curves are calculated intersect the area in random directions. In the general case, due to the velocity dispersion of PBHs within the cluster, the duration of a light curve may either increase or decrease. For the considered black hole cluster models, the velocity dispersion is significantly less than 0.1 of the cluster's own velocity in the Galactic halo; therefore, the resulting differences in the light curves are negligible. Furthermore, \noplps events differ from \plps ones by their complex shape, which is asymmetric relative to the peak, and this distinction persists even when the proper motion of the black holes is taken into account. The microlensing parameters are summarized in Table~\ref{tbl:microlens}.
\begin{table}
	\caption{Microlensing parameters for the PBH cluster models}
	\begin{tabular}{l}
		\hline
		\hline
		Distance to the cluster: $D_l =25$\,kpc \\
		Distance to the background stars: $D_s =50$\,kpc \\
		Cluster velocity as a whole: 200\,km/s \\
		Simulation area size in \mlens:  $2R_e(M_\odot)$\\
		Light curve duration:  $\sim 350^d$ \\
		Solid angle ratio \\  (simulation area/cluster) :  $4 \times10^{-8}$\\
		Simulation area resolution: $1000 \times 1000$\,pixels\\
		\hline
	\end{tabular}
	\label{tbl:microlens}
\end{table}

This study does not consider strong lensing by the PBH cluster itself, which occurs within a region of size \(l_{e}\) defined by the condition \(\Sigma (l_{e})=\Sigma _{cr}\). For the investigated PBH cluster models, the linear (\(l_{e}\)) and angular (\(\theta _{e}=l_{e}/D_{l}\)) scales of this region are provided in Table~\ref{tbl:models}. The largest angular scale is produced by model M1, with \(\theta _{e}=1.4\times 10^{-2\prime \prime }\), which is below the angular resolution of the leading space and ground-based telescopes, such as HST (\(0.1^{\prime \prime }\)), JWST (\(0.031^{\prime \prime }\)), and LSST (\(0.7^{\prime \prime }\)).

For each simulation area, sets of 50 to 150 magnification maps were generated, featuring a random collection of point-like PBH lenses with a uniform spatial distribution for the given values of \(\kappa \) and \(\gamma \). The mass distribution of these lenses follows the power law and the mass range specified in Table~\ref{tbl:params}. For low \(\kappa \) values at the cluster periphery, the average number of lenses per area was less than one on average. In such cases, a single lens was artificially placed within the area, subjected only to perturbations from the cluster as a whole, as defined by the external shear \(\gamma \).

\begin{figure*}[h] 
	\centering
	\includegraphics[width=0.99\textwidth]{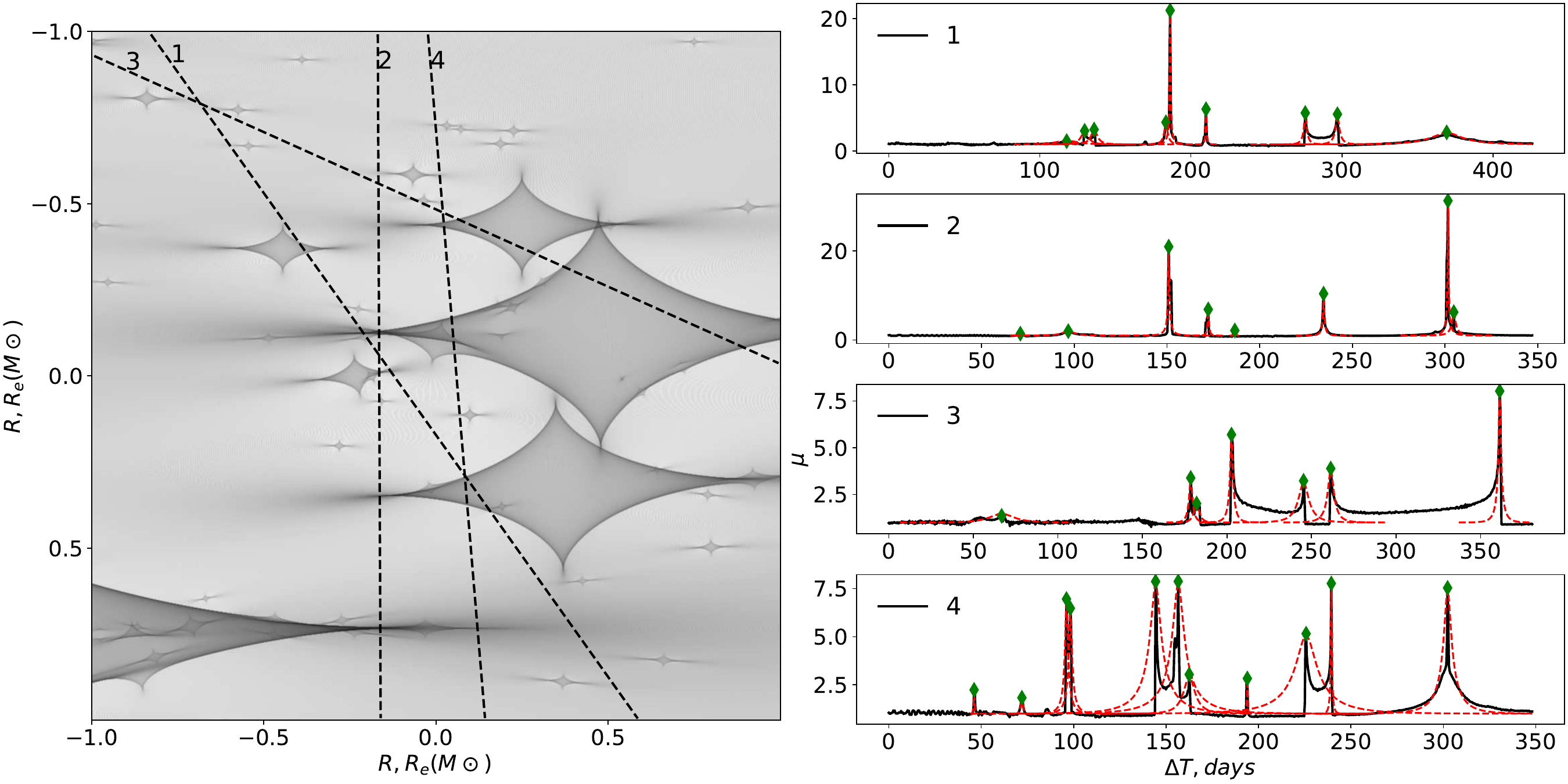}
	\caption{Illustration of the magnification map at $r = 0.014$\,pc from the cluster center (for model M1) with four sample trajectories of background sources (left). Light curves of the background sources along the corresponding trajectories featuring microlensing events (right). The microlensing maxima (peaks) are marked with green diamonds. The results of fitting the light curve peaks with the \plps profile are shown by red dashed lines.}
	\label{MapLc1}
\end{figure*}

\begin{figure*}[h] 
	\centering
	\begin{subfigure}[b]{0.49\textwidth}
		\includegraphics[width=\linewidth]{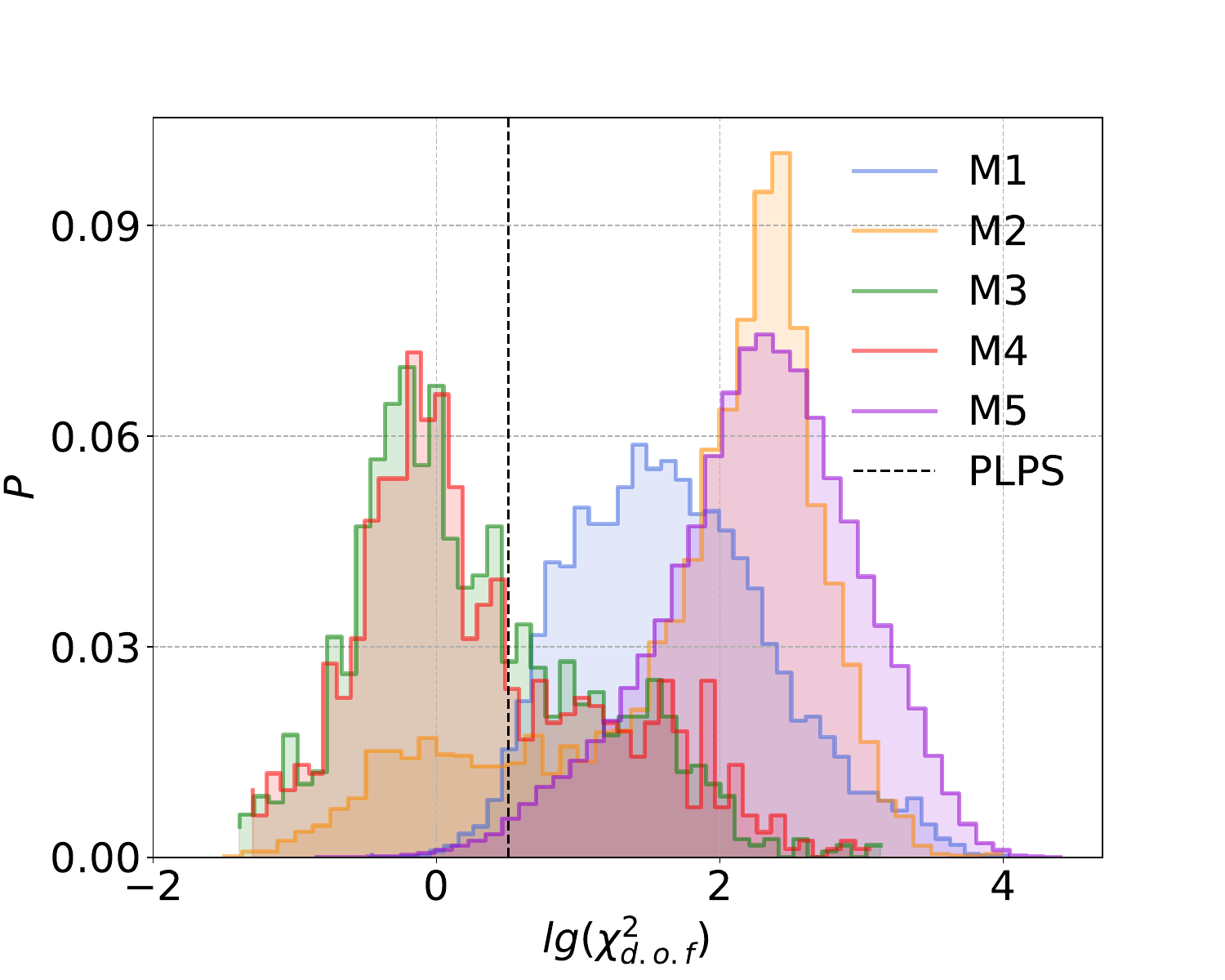}	
		\caption{$r = 0.01$\,pc}
		\label{Xi009}
	\end{subfigure}
	\hfill
	\begin{subfigure}[b]{0.49\textwidth}
		\includegraphics[width=\linewidth]{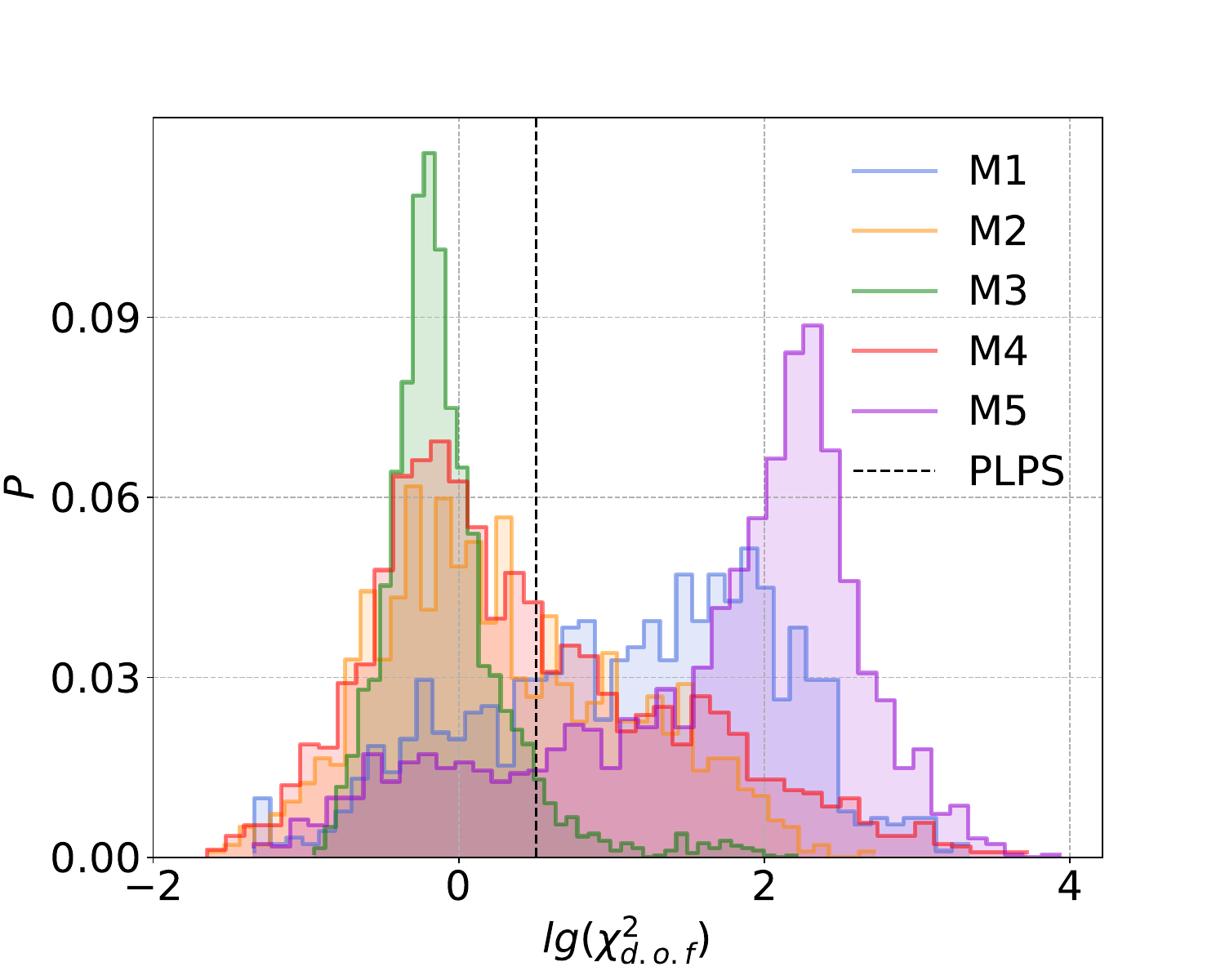}
		\caption{$r = 0.03$\,pc}		
		\label{Xi033}
	\end{subfigure}
	\hfill
	
	\begin{subfigure}{0.49\textwidth}
		\includegraphics[width=\linewidth]{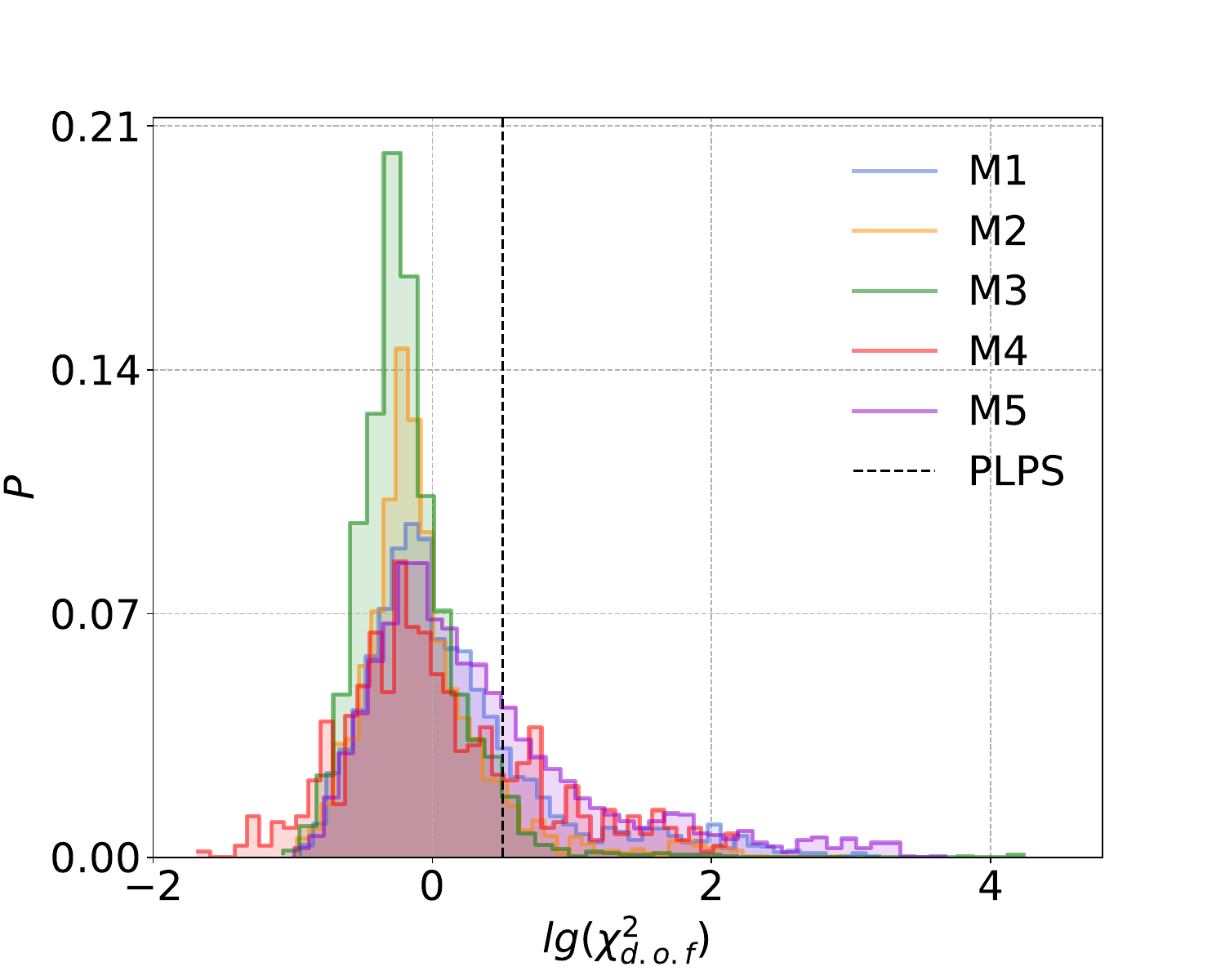}
		\caption{$r = 0.1$\,pc}
		\label{Xi100}
	\end{subfigure}
	\hfill
	\begin{subfigure}{0.49\textwidth}
		\includegraphics[width=\linewidth]{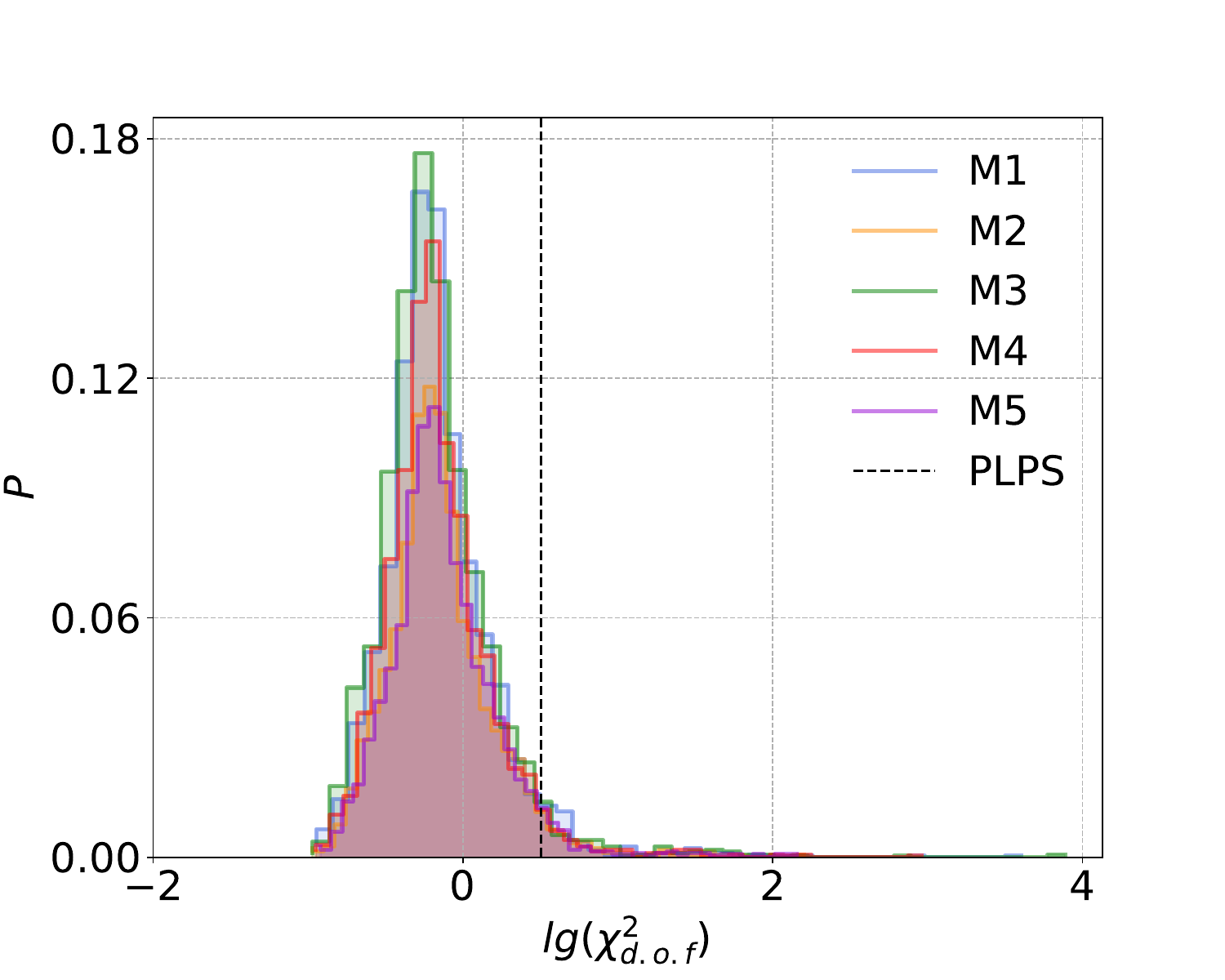}
		\caption{$r = 0.7$\,pc}
		\label{Xi700}
	\end{subfigure}
	\caption{The radial dependence of the $\chi^2_{d.o.f}$ goodness-of-fit criterion for the PBH cluster models. The black dashed line at $\chilim = 3.2$ separates \plps events (to the left) from \noplps events (to the right). As the distance from the cluster center increases, the fraction of \noplps events decreases}
	\label{Xi_stat}
\end{figure*}

It can be hypothesized that the evolution of the PBH cluster density profile is similar to that of globular clusters. Therefore, we additionally investigate two models utilizing the Plummer density profile, which is widely applied to describe globular clusters:
\begin{align} \label{eq:plammer}
	\rho(r) &= \frac{3M_{tot}}{4\pi R_{c}^3} \left( 1 + \frac{r^2}{R_{c}^2} \right)^{-\frac{5}{2}}
\end{align}

Each magnification map has a size of \(1000\times 1000\) pixels and covers an area occupying \(4\times 10^{-8}\) of the PBH cluster's solid angle.

Light curves were generated from the magnification map sets by convolving the map with the motion of a background source across it. In this study, the source size was chosen not to exceed 1 pixel on the model magnification maps, which corresponds to a background star size of \(<9R_{\odot }\) at a distance \(D_{s}\). The motion of the background source was assumed to be rectilinear with a random orientation to account for the asymmetric positioning relative to the cluster center. For each magnification map, at least 200 light curves were constructed to ensure sufficient statistical data.

\section{LIGHT CURVE ANALYSIS}\label{easyLC}

On each light curve, a search for microlensing events (peaks) was performed for which the magnification exceeds \(\mu >1.34\). This threshold corresponds to the magnification of a \plps model with an impact parameter equal to the Einstein-Chwolson radius. Each peak was fitted with a \plps profile using the \(SciPy\) library \citep{virtanen2020scipy}. A demonstration of the algorithm's performance is shown in Fig.~\ref{MapLc1}. The left panel displays the magnification map with four examples of background source trajectories, while the right panel shows the corresponding light curves. The detected peaks are marked in green, and the \plps profiles used to fit these peaks are shown in red.

In microlensing studies, the $\chi^2_{d.o.f}$ goodness-of-fit criterion has been employed to detect \plps microlensing events (Equation \ref{form_Xi}):
\begin{align}
	\chi^2_{d.o.f} = \frac{1}{N_{d.o.f}}\sum_i \left( \frac{m_{obs, i} - m_{fit, i}}{\sigma_i}\right)^2
	\label{form_Xi}
\end{align}
, where \(m_{obs,i}\) is the observed magnitude of the light curve, \(m_{fit,i}\) is the magnitude of the \plps fit (\ref{eq:One_lens_magnification}), \(N_{d.o.f.}\) is the number of degrees of freedom, and \(\sigma _{i}\) is the photometric error. This \(\chi _{d.o.f.}^{2}\) criterion was applied to distinguish between \plps and \noplps events on the simulated light curves, where \(m_{obs,i}\) was replaced by \(m_{mod,i}=-2.5\lg (\mu )\) — the magnification values \(\mu \) converted into stellar magnitudes. The observational photometric error was simulated as \(\sigma _{i}=0.05\times 10^{0.2m_{mod,i}}\) according to the data from \cite{1995ApJ...440...22M}.

During the fitting procedure, light curve events were classified as \plps if the criterion was $\chi^2_{d.o.f} < \chilim$; otherwise, they were categorized as \noplps. The threshold value \chilim was derived from an analysis of magnification maps generated for single point lenses, corresponding to \plps events. More than 15,000 light curves were constructed for point lenses of various masses. The $\chi^2_{d.o.f}$ values were calculated for all single-lens microlensing events, and the threshold \chilim was determined such that $95\%$ of these events were classified as \plps. Figure \ref{Xi_stat} shows the $\chi^2_{d.o.f}$ distributions as a function of radius $r$ , with the threshold value \chilim indicated. From these distributions, the event ratio is determined as $\fplps(r) = N_{PLPS}(r)\,/\, N_{tot}(r)$, where $N_{PLPS}(r)$ is the number of \plps events and $N_{tot}(r) = N_{PLPS}(r) + N_{no-PLPS}(r)$ is the total number of events.

Power-law density profile models M1–M3 exhibit a decrease in \fplps as the index \(\alpha \) increases, which is evident in Fig. \ref{PercentEasy} for regions located near the PBH cluster center. Simultaneously, for all cluster models, \fplps tends toward 1 as the radius \(r\) increases from the center to the periphery; this is expected, since the lens density, their mutual influence, and the impact of the cluster center all diminish toward the edge. For the Plummer profile models M4 and M5, Fig. \ref{PercentEasy} shows a drop in \fplps as the core radius \(R_{c}\) decreases; this occurs because the majority of lenses are concentrated within a more compact core, where mutual interactions cause their corresponding caustics to deviate from the \plps case. In the central region of M5, \plps events vanish completely. Model M1, which features the steepest power-law density decline, demonstrates similar behavior.

\noplps events are not recorded in observations; consequently, the dark matter mass associated with them is not included in the microlensing constraints on $f_{PBH}$. The fraction of unaccounted-for matter \gnoplps in the form of PBHs is calculated by convolving the surface density $\Sigma(r)$ with the event fraction $\fplps(r)$ using formula (\ref{fractiom_simple}):
\begin{align}
	\begin{split}
		\gnoplps &= 1 - \frac{M_{PLPS}}{M_{tot}} = \\
		&= 1 - \frac{2 \pi}{M_{tot}} \int_0^{R_{cl}}{  \Sigma(r)  \fplps(r) rdr},
	\end{split}
	\label{fractiom_simple}
\end{align}
where $M_{PLPS}$ is the mass responsible for \plps events and $M_{tot}$ is the total mass.

The values of \gnoplps for the PBH cluster models are presented in Table \ref{tbl:hard_event}.
\begin{table}
	\caption{Mass fraction of the cluster \gnoplps producing microlensing \noplps events.}
	\begin{center}
		\begin{tabular}{l c}
			\hline
			\hline
			Model  & \gnoplps \\
			\hline
			M1 $(\alpha = 2.8)$ & 0.60 \\
			M2 $(\alpha = 2.5)$ & 0.22 \\
			M3 $(\alpha = 2.0)$ & 0.11 \\
			M4 $(R_{core} = 10^{-1}$ pc) & 0.20 \\
			M5 $(R_{core} = 10^{-2}$ pc) & 0.93 \\
			\hline		
		\end{tabular}
	\end{center}
	\label{tbl:hard_event}
\end{table}

The maximum fraction of mass hidden from observational experiments reaches $0.93$ for model M5. In the initial cluster model M1, 60\% of the mass is "hidden" from observations; however, this is insufficient to completely lift the constraints on the PBH fraction in dark matter. If the density profile of a PBH cluster is less "steep," as in models M2 and M3, the fraction of unaccounted-for mass \gnoplps decreases. In this case, the constraints from MACHO microlensing observations extend to such clusters, making their existence in nature unlikely. Conversely, if PBH clusters become more concentrated during their evolution, the opposite effect occurs, and the fraction of unaccounted-for mass \gnoplps increases.

\section{CONCLUSIONS}\label{conclusions}

This study examines the effect of gravitational microlensing on primordial black hole (PBH) clusters to investigate the fraction of matter that may remain 'hidden' from microlensing observations employing the point-lens point-source (PLPS) model for MACHO detection. We evaluate the impact of the PBH cluster density profile on the light curves of background sources, utilizing five distinct models with power-law and Plummer density profiles.

Magnification maps and corresponding light curves for the cluster models were generated using the ray-tracing method. Microlensing events identified in the simulated light curves were fitted with the PLPS model. Based on the goodness-of-fit criterion, these events were categorized into two groups: PLPS and non-PLPS. Statistical analysis reveals that a significant fraction of the events is inconsistent with the PLPS model. Specifically, for the M5 model—which features the highest central density—the mass of the PBH cluster producing non-PLPS events can reach up to 93\%. 

However, in the peripheral regions of all studied clusters, the microlensing events remain indistinguishable from the PLPS model; thus, a substantial fraction of PBHs continues to act as isolated lenses. Consequently, the clustering of PBHs does not fully eliminate the microlensing constraints on the PBH contribution to dark matter.

It is well known that in the linear approximation, positioning the lens at the midpoint of the line of sight ($D_{l}=\frac{1}{2}D_{s}$) maximizes the optical depth of the cluster for a given $D_{l}$. For other lens positions, the influence of PBH clustering on microlensing events diminishes, leading to a decrease in the fraction of 'hidden' dark matter mass, \gnoplps.

The deployment of next-generation wide-field telescopes, such as the Vera C. Rubin Observatory, will enable the detection of microlensing events towards stars distributed throughout the Galactic volume \citep{Sajadian2019}. Since the average distance to Galactic stars is smaller than the distance to the LMC, the values of \(\kappa \) and \(\gamma \) for a given PBH cluster are reduced. Consequently, the mutual interference between lenses that leads to non-PLPS events diminishes, resulting in a lower fraction of 'hidden' dark mass, \gnoplps. It is evident that while clustering PBHs can relax the microlensing constraints on $f_{PBH}$ , it does not entirely eliminate them

\section*{ACKNOWLEDGMENTS}
We are grateful to J. Wambsganss for the development of the \mlens code and for the permission to use it to model the microlensing. 
We thank E.M. Urvachev, A.V. Yudin, E.I. Sorokina, and N.S. Lyskov for their interest in the paper and useful discussions.

\section*{FUNDING}
The modeling of the light curves for microlensing events and their statistical analysis performed by P.V. Baklanov were supported by RSF grant no. 23-12-00220. 
The modeling of PBH clusters performed by K.M. Belotsky was supported by RSF grant no. 23-42-00066.



\bibliographystyle{apalike}
\bibliography{references}

@ARTICLE{Byalko1969orig,
	author = {{Byalko}, A.~V.},
	title = "{Focusing of Radiation by a Gravitational Field.}",
	journal = {\azh},
	year = 1969,
	month = jan,
	volume = {46},
	pages = {998},
	adsurl = {https://ui.adsabs.harvard.edu/abs/1969AZh....46..998B}
}

@ARTICLE{1986MNRAS.219..333S,
	author = {{Subramanian}, K. and {Cowling}, S.~A.},
	title = "{On local conditions for multiple imaging by bounded, smooth gravitational lenses}",
	journal = {\mnras},
	year = 1986,
	month = mar,
	volume = {219},
	pages = {333-346},
	doi = {10.1093/mnras/219.2.333},
	adsurl = {https://ui.adsabs.harvard.edu/abs/1986MNRAS.219..333S}
}

@ARTICLE{hawking1971gravitationally,
	author={{Hawking}, S.},
	title={Gravitationally collapsed objects of very low mass},  
	journal={\mnras},
	volume={152},
	number={1},
	pages={75--78},
	year={1971},
	publisher={Oxford University Press}
}

@ARTICLE{Zeldovich1966,
	author = {{Zeldovich}, Ya. B. and {Novikov}, I.~D.},
	title = "{The Hypothesis of Cores Retarded during Expansion and the Hot Cosmological Model}",
	journal = {\azh},
	year = 1966,
	month = jan,
	volume = {43},
	pages = {758},
	adsurl = {https://ui.adsabs.harvard.edu/abs/1966AZh....43..758Z}
}

@article{Griest_2014,
	doi = {10.1088/0004-637X/786/2/158},
	url = {https://dx.doi.org/10.1088/0004-637X/786/2/158},
	year = {2014},
	month = {apr},
	publisher = {The American Astronomical Society},
	volume = {786},
	number = {2},
	pages = {158},
	author = {Griest, Kim and Cieplak, Agnieszka M. and Lehner, Matthew J.},
	title = {EXPERIMENTAL LIMITS ON PRIMORDIAL BLACK HOLE DARK MATTER FROM THE FIRST 2 YR OF KEPLER DATA},
	journal = {The Astrophysical Journal}
}

@article{PhysRevLett.111.181302,
	title = {New Limits on Primordial Black Hole Dark Matter from an Analysis of Kepler Source Microlensing Data},
	author = {Griest, Kim and Cieplak, Agnieszka M. and Lehner, Matthew J.},
	journal = {Phys. Rev. Lett.},
	volume = {111},
	issue = {18},
	pages = {181302},
	numpages = {5},
	year = {2013},
	month = {Oct},
	publisher = {American Physical Society},
	doi = {10.1103/PhysRevLett.111.181302},
	url = {https://link.aps.org/doi/10.1103/PhysRevLett.111.181302}
}

@article{petavc2022microlensing,
	title={Microlensing constraints on clustered primordial black holes},
	author={Peta{\v{c}}, Mihael and Lavalle, Julien and Jedamzik, Karsten},
	journal={Physical Review D},
	volume={105},
	number={8},
	pages={083520},
	year={2022},
	publisher={APS}
}

@article{GARCIABELLIDO2018144,
	title = {Constraints from microlensing experiments on clustered primordial black holes},
	journal = {Physics of the Dark Universe},
	year = {2018},
	author = {Juan García-Bellido and Sébastien Clesse},
}

@article{gorton2022effect,
	title={Effect of clustering on primordial black hole microlensing constraints},
	author={Gorton, Matthew and Green, Anne M},
	journal={Journal of Cosmology and Astroparticle Physics},
	year={2022},
}

@article{carr2021constraints,
	title={Constraints on primordial black holes},
	author={Carr, Bernard and others},
	journal={Reports on Progress in Physics},
	year={2021},
}

@ARTICLE{2000ApJ...541..270A,
	author = {{Alcock}, C. and others},
	title = "{Binary Microlensing Events from the MACHO Project}",
	journal = {\apj},
	year = 2000,
	volume = {541},
	pages = {270},
}

@article{Paczynski:1985jf,
	author = "Paczynski, Bohdan",
	title = "{Gravitational microlensing by the galactic halo}",
	doi = "10.1086/164140",
	journal = "Astrophys. J.",
	volume = "304",
	pages = "1--5",
	year = "1986"
}

@ARTICLE{1995ApJ...440...22M,
	author = {{Mao}, S. and {Di Stefano}, R.},
	title = "{Interpretation of Gravitational Microlensing by Binary Systems}",
	journal = {\apj},
	year = 1995,
	month = feb,
	volume = {440},
	pages = {22},
	doi = {10.1086/175244},
}

@ARTICLE{2019PhRvD..99h3503N,
	author = {{Niikura}, Hiroko and others},
	title = "{Constraints on Earth-mass primordial black holes from OGLE 5-year microlensing events}",
	journal = {\prd},
	year = 2019,
	month = apr,
	volume = {99},
	number = {8},
	pages = {083503},
	doi = {10.1103/PhysRevD.99.083503},
}

@ARTICLE{2019NatAs...3..524N,
	author = {{Niikura}, Hiroko and others},
	title = "{Microlensing constraints on primordial black holes with Subaru/HSC Andromeda observations}",
	journal = {Nature Astronomy},
	year = 2019,
	month = apr,
	volume = {3},
	pages = {524-534},
	doi = {10.1038/s41550-019-0723-1},
}

@article{refId0,
	author = {{Calchi Novati, S.} and others},
	title = {POINT-AGAPE pixel lensing survey of M31 - Evidence for a MACHO contribution to galactic halos},
	journal = {\aap},
	year = 2005,
	volume = 443,
	number = 3,
	pages = "911-928",
}

@article{WAMBSGANSS1999353,
	title = {Gravitational lensing: numerical simulations with a hierarchical tree code},
	journal = {Journal of Computational and Applied Mathematics},
	volume = {109},
	number = {1},
	pages = {353-372},
	year = {1999},
	author = {Joachim Wambsganss},
}

@ARTICLE{2005GrCo...11...99D,
	author = {{Dokuchaev}, V.~I. and {Eroshenko}, Yu. N. and {Rubin}, S.~G.},
	title = "{Quasars formation around clusters of primordial black holes}",
	journal = {Gravitation and Cosmology},
	year = 2005,
	month = jun,
	volume = {11},
	pages = {99-104},
	doi = {10.108550/arXiv.astro-ph/0412418},
}

@ARTICLE{2001JETP...92..921R,
	author = {{Rubin}, S.~G. and {Sakharov}, A.~S. and {Khlopov}, M. Yu.},
	title = "{The Formation of Primary Galactic Nuclei during Phase Transitions in the Early Universe}",
	journal = {Soviet Journal of Experimental and Theoretical Physics},
	year = 2001,
	month = jun,
	volume = {92},
	number = {6},
	pages = {921-929},
	doi = {10.1134/1.1385631},
}

@ARTICLE{2000hep.ph....5271R,
	author = {{Rubin}, S.~G. and {Khlopov}, M. Yu. and {Sakharov}, A.~S.},
	title = "{Primordial Black Holes from Non-Equilibrium Second Order Phase Transition}",
	journal = {arXiv e-prints},
	year = 2000,
	month = may,
	pages = {hep-ph/0005271},
	doi = {10.48550/arXiv.hep-ph/0005271},
}

@ARTICLE{2020Univ....6..158B,
	author = {{Berezin}, Victor and others},
	title = "{Formation and Clustering of Primordial Black Holes in Brans-Dicke Theory}",
	journal = {Universe},
	year = 2020,
	month = sep,
	volume = {6},
	number = {10},
	pages = {158},
	doi = {10.3390/universe6100158},
}

@ARTICLE{young2020,
	author = {{Young}, Sam and {Byrnes}, Christian T.},
	title = "{Initial clustering and the primordial black hole merger rate}",
	journal = {\jcap},
	year = 2020,
	month = mar,
	volume = {2020},
	number = {3},
	pages = {004},
	doi = {10.1088/1475-7516/2020/03/004},
}

@ARTICLE{2019PhRvD.100l3544M,
	author = {{Matsubara}, Takahiko and others},
	title = "{Clustering of primordial black holes formed in a matter-dominated epoch}",
	journal = {\prd},
	year = 2019,
	month = dec,
	volume = {100},
	number = {12},
	pages = {123544},
	doi = {10.1103/PhysRevD.100.123544},
}

@article{Young_2015,
	year = {2015},
	month = {apr},
	volume = {2015},
	number = {04},
	pages = {034},
	author = {Sam Young and Christian T. Byrnes},
	title = {Signatures of non-gaussianity in the isocurvature modes of primordial black hole dark matter},
	journal = {Journal of Cosmology and Astroparticle Physics},
	doi = {10.1088/1475-7516/2015/04/034},
}

@ARTICLE{2015PhRvD..91l3534T,
	author = {{Tada}, Yuichiro and {Yokoyama}, Shuichiro},
	title = "{Primordial black holes as biased tracers}",
	journal = {\prd},
	year = 2015,
	month = jun,
	volume = {91},
	number = {12},
	pages = {123534},
	doi = {10.1103/PhysRevD.91.123534},
}

@article{10.1093/ptep/ptz105,
	author = {Suyama, Teruaki and Yokoyama, Shuichiro},
	title = "{Clustering of primordial black holes with non-Gaussian initial fluctuations}",
	journal = {Progress of Theoretical and Experimental Physics},
	volume = {2019},
	number = {10},
	year = {2019},
	month = {10},
	doi = {10.1093/ptep/ptz105},
}

@article{Atal_2020,
	year = {2020},
	month = {nov},
	volume = {2020},
	number = {11},
	pages = {036},
	author = {Vicente Atal and others},
	title = {The role of non-Gaussianity in the clustering of primordial black holes},
	journal = {Journal of Cosmology and Astroparticle Physics},
	doi = {10.1088/1475-7516/2020/11/036},
}

@ARTICLE{2020JCAP...11..036A,
	author = {{Atal}, Vicente and {Sanglas}, Albert and {Triantafyllou}, Nikolaos},
	title = "{LIGO/Virgo black holes and dark matter: the effect of spatial clustering}",
	journal = {\jcap},
	year = 2020,
	month = nov,
	volume = {2020},
	number = {11},
	pages = {036},
	doi = {10.1088/1475-7516/2020/11/036},
}

@ARTICLE{2021Univ....7...18T,
	author = {{Trashorras}, Manuel and {Garc{\'\i}a-Bellido}, Juan and {Nesseris}, Savvas},
	title = "{The Clustering Dynamics of Primordial Black Boles in N-Body Simulations}",
	journal = {Universe},
	year = 2021,
	month = jan,
	volume = {7},
	number = {1},
	pages = {18},
	doi = {10.3390/universe7010018},
}

@article{korol2020merger,
	title={Merger rates in primordial black hole clusters without initial binaries},
	author={Korol, Valeriya and others},
	journal={\mnras},
	volume={496},
	number={1},
	pages={994--1000},
	year={2020},
	doi = {10.1093/mnras/staa1644},
}

@article{Belotsky:2018wph,
	author         = {Belotsky, Konstantin M. and others},
	title          = "{Clusters of primordial black holes}",
	journal        = "Eur. Phys. J. C",
	volume         = "79",
	year           = "2019",
	number         = "3",
	pages          = "246",
	doi            = "10.1140/epjc/s10052-019-6741-4",
}

@article{Khlopov:2004sc,
	author= {Khlopov, M.Yu. and Rubin, S.G. and Sakharov, A.S.},
	title= {Primordial structure of massive black hole clusters},
	journal        = "Astropart. Phys.",
	volume         = "23",
	year           = "2005",
	pages          = "265",
	doi            = "10.1016/j.astropartphys.2004.12.002",
}

@article{Schneider1992,
	author = {Schneider, Peter and Ehlers, J{\"{u}}rgen and Falco, Emilio E.},
	title = {{Gravitational Lenses}},
	journal = {Springer Berlin Heidelberg},
	year = {1992},
	doi = {10.1007/978-3-662-03758-4},
}

@article{MACHO2000,
	title={The MACHO Project: Microlensing Results from 5.7 Years of Large Magellanic Cloud Observations},
	author={Alcock, C. and others},
	journal={The Astrophysical Journal},
	year={2000},
	volume={542},
	pages={281–307},
	doi={10.1086/309512}
}

@ARTICLE{2007A&A...469..387T,
	author = {{Tisserand}, P. and others},
	title = "{Limits on the Macho content of the Galactic Halo from the EROS-2 Survey of the Magellanic Clouds}",
	journal = {\aap},
	year = 2007,
	volume = {469},
	pages = {387-404},
	doi = {10.1051/0004-6361:20066017}
}

@article{virtanen2020scipy,
	title={SciPy 1.0: fundamental algorithms for scientific computing in Python},
	author={Virtanen, Pauli and others},
	journal={Nature methods},
	volume={17},
	year={2020},
	pages={261--272}
}

@article{mroz2025,
	title={Microlensing Optical Depth, Event Rate, and Limits on Compact Objects in Dark Matter Based on 20 Yr of OGLE Observations of the Small Magellanic Cloud},
	author={Mr{\'o}z, Przemek and others},
	journal={The Astrophysical Journal Supplement Series},
	volume={280},
	year={2025},
	pages={49}
}

@article{Sajadian2019,
	title={Predictions for the Detection and Characterization of Galactic Disk Microlensing Events by LSST},
	author={Sajadian, Sedighe and Poleski, Rados{\l}aw},
	journal={The Astrophysical Journal},
	volume={871},
	year={2019},
	pages={205}
}

@ARTICLE{Dolgov1993,
	author = {{Dolgov}, Alexandre and {Silk}, Joseph},
	title = "{Baryon isocurvature fluctuations at small scales and baryonic dark matter}",
	journal = {\prd},
	year = 1993,
	volume = {47},
	pages = {4244-4255},
	doi = {10.1103/PhysRevD.47.4244}
}

@article{Dolgov2018,
	author = {A. D. Dolgov},
	title = {Massive and supermassive black holes in the contemporary and early Universe and problems in cosmology and astrophysics},
	year = {2018},
	journal = {Usp. Fiz. Nauk},
	volume = {188},
	pages = {121-142},
	doi = {10.3367/UFNr.2017.06.038153}
}

\end{document}